 \documentclass[final,5p,times,twocolumn,authoryear]{elsarticle}

\usepackage{amssymb}
\usepackage{amsmath}
\usepackage{lipsum}
\usepackage{natbib}
\usepackage[colorlinks=true, citecolor=blue, linkcolor=blue]{hyperref}

\newcommand{\fief}{\Phi_{\rm eff}}
\newcommand{\mb}{M_{\rm BH}}
\newcommand{\rg}{r_{\rm g}}
\newcommand{\tg}{t_{\rm g}}
\newcommand{\ak}{a_{\rm k}}
\newcommand{\rc}{r_{\rm c}}

\def\lsim{\lower.5ex\hbox{$\; \buildrel < \over \sim \;$}}
\def\gsim{\lower.5ex\hbox{$\; \buildrel > \over \sim \;$}}
\def\simeq{\lower.3ex\hbox{$\; \buildrel \sim \over - \;$}}
\def\lou{\lambda_{\rm ou}}
\def\vo{v_{\rm ou}}
\def\tho{\Theta_{\rm ou}}
\def\ro{r_{\rm ou}}
\def\alfc{\alpha_{\rm cr}}
\def\rs{r_{\rm sh}}
\journal{High Energy Astrophysics}

\begin{document}

\begin{frontmatter}



\title{QPOs from the Viscous Transonic Accretion Flow Around a Spinning Black Hole}


\author[1,2]{Sanjit Debnath}
\author[1]{Indranil Chattopadhyay}
\author[1,3]{Soumyadip Mandal}
\author[4]{Raj Kishor Joshi}
\author[1,2]{Priyesh Kumar Tripathi}
\author[2]{M. Saleem Khan}

\address[1]{Aryabhatta Research Institute of Observational Sciences (ARIES), Manora Peak, Nainital, 263001, India}
\address[2]{Department of Applied Physics, Mahatma Jyotiba Phule Rohilkhand University, Bareilly, Uttar Pradesh, 243006, India}
\address[3]{Indian Institute of Technology Roorkee, Roorkee, Uttarakhand, 247667, India}
\address[4]{Nicolaus Copernicus Astronomical Center, Polish Academy of Sciences, Bartycka 18, PL-00-716, Warsaw, Poland}

\begin{abstract}
We investigate the dynamics of transonic advective accretion flows around spinning black holes in the presence of viscosity. The spacetime of a Kerr black hole is approximated using a pseudo-potential. We study viscously driven shock oscillations over a range of black hole spin parameters. Our results show that the frequency range of quasi-periodic oscillations (QPOs) obtained from the power density spectra depends strongly on the black hole spin. Low-spin systems predominantly exhibit low-frequency QPOs, whereas rapidly rotating black holes ($\gtrsim 0.9$) produce QPOs spanning a broad range from low to high frequencies, comparable to those observed in black hole X-ray binaries. We further obtain a correlation between the QPO frequency and the power-law photon index ($\Gamma$–$\nu_{\rm QPO}$) by computing the spectrum for a $10M_\odot$ black hole.
\end{abstract}




\end{frontmatter}




\section{Introduction}
\label{introduction}
Black holes, spanning from stellar mass to supermassive scales, are generally observed through the radiation emitted from the matter accreting onto them. This accreting matter forms a disk whose structure and emission properties are governed by complex fluid dynamics and radiative processes. Recent black hole (BH) shadow images, such as those of Sgr A* and M87  \citep{2019ApJ...875L...1E,2021ApJ...910L..13E,2022ApJ...930L..12E}, reinforce the idea that the swirling hot plasma around the black hole is responsible for the observed radiation from these sources. The continuum emission from black hole X-ray binaries (BHXRBs, or microquasars) is dominated by X-rays and is related to accretion activity, while radiation at lower frequencies, such as IR or radio, is associated with outflows or jets.
BHXRBs are commonly observed to transit between various spectral states \citep[hard, soft, and intermediate, see][]{2004MNRAS.355.1105F,2019NewAR..8501524I}. These systems exhibit a range of disk behaviors, including state transitions and =variability, radio jets, and quasi-periodic oscillations (QPOs). A striking timing feature is the occurrence of QPOs over a wide frequency range, especially pronounced in the hard and intermediate states. QPOs in BHXRBs are classified into two types: low-frequency QPOs or LFQPOs and high-frequency QPOs or HFQPOs \citep[][and references therein]{2002ApJ...572..392B}.

Many studies of accretion onto compact objects and emission from systems containing BHs reveal a pronounced power-law component in their spectral energy density (SED), prompting the development of numerous accretion models in the 1970s. The most widely accepted accretion disk model, known as the Keplerian disk or Shakura-Sunyaev disk \citep[KD;][]{1973A&A....24..337S,1973blho.conf..343N}, accounts only for the thermal multicolor blackbody emission and not the power-law contribution to the SED. Consequently, many alternative models were proposed to address the power-law components, all featuring non-Keplerian angular momentum distributions \citep{1977ApJ...214..840I,1980A&A....88...23P,1980ApJ...240..271L,1987PASJ...39..309F,1989ApJ...347..365C,1994ApJ...428L..13N}.

{Among} these models, the transonic accretion flow \citep{1980ApJ...240..271L} {permits} a variety of solutions, {such as flows in which matter plunges into a black hole through a single sonic point ($\rc$, where the infall speed matches the local sound speed), or solutions featuring two sonic points linked by a shock} \citep{1987PASJ...39..309F,1989ApJ...347..365C}. This scenario allows multiple {pathways for matter to enter the BH}. Notably, the post-shock disk (PSD) can serve as a reservoir of hot electrons, {the hypothesized} Compton cloud, potentially producing the inverse-Comptonized power-law portion of the SED \citep{1995ApJ...455..623C,2005IJMPD..14..933M,2006ApJ...642L..49C,2020A&A...642A.209S,2022JApA...43...34S}. 
Additionally, \cite{1996ApJ...457..805M} showed {that the hotter and denser PSD  is more susceptible to radiative cooling than the pre-shock region}. This would cause the shock front to move closer to the BH, compressing the PSD. Once resonance is reached, the shock front would oscillate. Oscillations in the PSD cause the power-law photons to oscillate, exactly as in a QPO. Many papers using numerical simulations also presented various factors {that} may destabilize the post-shock dynamics \citep{1995ApJ...452..364R,2010MNRAS.403..516G,2012MNRAS.425.2413O,2012ApJ...758..114G,2013MNRAS.430.2836G,2014MNRAS.437.1329G,2015MNRAS.448.3221G,2017MNRAS.472.4327S,2017MNRAS.472..542K,2019MNRAS.482.3636K, 2020ApJ...904...21P,2022MNRAS.514.5074O,2023A&A...678A.141O,2023MNRAS.519.4550G,2025ApJ...990...35D,2025JHEAp..4700395Z,2025ApJ...990...12M,2025RAA....25b5013H,2026ApJ...997..277D,2026Univ...12...77J}. More specifically, 
\cite{1998MNRAS.299..799L,2014MNRAS.442..251D,2011ApJ...728..142L,2012MNRAS.421..666G,2015MNRAS.453..147O,2016ApJ...831...33L,2024MNRAS.528.3964D,2025ApJ...994...48D}
showed {how} viscosity can also trigger shock oscillation by transporting angular momentum at different rates in the inner disk {as} compared to that in the outer disk. However, all theoretical and simulation work reported only low-frequency QPOs (sub-Hz to around 10 Hz for a 10 solar mass BH), or LFQPOs.
Although {few papers theoretically estimated  HFQPO \citep{2009MNRAS.396.1038M,2017MNRAS.471.4806A}}.

In this work, we investigate the role of viscosity in regulating shock oscillations in accretion flows around spinning black holes using a time-dependent 1D hydrodynamic framework. We adopt a relativistically corrected equation of state (EoS) as proposed by \cite{2009ApJ...694..492C} instead of the ideal fixed-$\gamma$ EoS. This EoS provides a more accurate description of the flow temperature and, consequently, a more consistent treatment of viscous angular momentum transport, as discussed in \cite{2024MNRAS.528.3964D}. For a given black hole spin, we explore a wide parameter space by systematically varying the viscosity parameter, considering different values of the specific angular momentum ($\lambda$) and specific energy ($\epsilon$) at the outer boundary. The flow is assumed to be a fully ionised electron–proton plasma. We examine the interplay between spin, viscosity, and shock dynamics, and their possible connection to the origin of QPOs. In particular, we determine the critical viscosity required to trigger shock oscillations for different spin values and analyse its dependence on the flow parameters (specific energy and angular momentum). We also investigate the influence of black hole spin on the resulting QPO frequencies spanning low to high frequency ranges. This provides a consistent framework to interpret the diversity of QPOs observed in black hole X-ray binaries.

The paper is organised as follows. In Section~\ref{sec:equatins}, we present the governing equations. Section~\ref{sec:code} describes the numerical method and boundary conditions. The results are presented in Section~\ref{results}, followed by the discussion and conclusions in Section~\ref{discussion}.

\section{Basic Equations}\label{sec:equatins}
\subsection{Hydrodynamics}
We have studied the transonic, viscous, low-angular-momentum accretion flow around a rotating black hole in spherical polar coordinates. We have adopted the geometrical unit system as $G = \mb = c =$ 1, where $G$ is
the universal gravitational constant, $\mb$ is the mass of the black hole, and $c$ is the speed of light. So the unit of length and time are $\rg=G \mb /c^2$, $\tg = \rg/c = \mb /c^3$, respectively. To mimic the gravity of a rotating black hole, we have selected a pseudo-potential given by \cite{2018PhRvD..98h3004D,2024IJMPD..3350059B} as,
\begin{equation}
\fief
= 1 + \frac{1}{2} \ln \left[
\frac{r\left(r^{2} - 2\rg r + \ak^{2}\right)}
     {\ak^{2}(r+2) - 4 \ak \lambda + r^{3} - \lambda^{2}(r-2)}
\right],
\label{eq:potential}
\end{equation}
where $\ak$ represents the spin of the black hole, $\lambda$ is the specific angular momentum of fluid. However, the effective potential contains contributions from both gravity and the centrifugal force. While this form is suitable for use in the radial momentum equation, it cannot be directly applied in the energy balance equation since the centrifugal force does no work. Therefore, to separate the gravitational part of the potential, we set $\lambda$=0 in the expression for the effective potential. The resulting potential, representing the contribution from gravity alone, is given by
\begin{equation}
\Phi_{\rm grav}
= 1 + \frac{1}{2} \ln \left[
\frac{r\left(r^{2} - 2\rg r + \ak^{2}\right)}
     {\ak^{2}(r+2)+ r^{3} }
\right].
\label{eq: grav_potential}
\end{equation}
This potential has been adopted in many accretion works \citep{2020MNRAS.496.3043D,2024MNRAS.527.1745A,2025JCAP...10..090J,2025JCAP...05..055S,2025ApJ...980...68K,2025ApJ...987..145A}.
We have considered only the $r-\phi$ component of viscous stress, as it transports the angular momentum. The compact form of the conserved equations of motion in spherical coordinates is given by:
\begin{equation}
 \frac{\partial \mathbf{q}}{\partial t} + \frac{1}{r^2} \frac{\partial (r^2\mathbf{F^r})}{\partial r} +\frac{1}{r \sin\theta} \frac{\partial(\sin\theta \mathbf{F^\theta})}{\partial \theta} + \frac{1}{r \sin\theta} \frac{\partial( \mathbf{F^\phi})}{\partial \phi}   = \mathbf{S}.
 \label{eq:conserve_full}
\end{equation}
The one dimensional ($\partial/\partial \theta \equiv \partial/\partial \phi \equiv 0$ \& $v_\theta = 0$), equation of motion is:
\begin{equation}
 \frac{\partial \mathbf{q}}{\partial t} + \frac{1}{r^2} \frac{\partial (r^2\mathbf{F^r})}{\partial r}  = \mathbf{S}.
 \label{eq:conserve}
\end{equation}
where the conserved variables ($\mathbf{q}$), corresponding fluxes ($\mathbf{F^r}$), the source ($\mathbf{S}$) and the primitive variables ($\mathbf{w}$) are given as follows:
\begin{equation}
\mathbf{q}= 
\begin{bmatrix}
\rho \\
 M_r \\
 M_{\theta} \\
  M_\lambda \\
E \\
\end{bmatrix}
=
\begin{bmatrix}
\rho \\
 \rho v_r \\
\rho v_{\theta} \\
 \rho \lambda \\
\rho v^2/2 +e \\
\end{bmatrix};
~~ \mathbf{w}= 
\begin{bmatrix}
\rho \\
 v_r \\
 v_{\theta} \\
  \lambda\\
 p \\
\end{bmatrix} 
\label{eq:state_primitv}
\end{equation}
\begin{equation}
\mathbf{F^r}= 
\begin{bmatrix}
\rho v_r \\
v_r M_r+p \\
v_r M_{\theta} \\
v_r M_\lambda \\
(E+p)v_r\\
\end{bmatrix}, 
\mathbf{S}= 
\begin{bmatrix}
0 \\
-\rho\frac{d\Phi_{\rm eff}}{dr}+\frac{2p}{r} \\
0 \\
 S_\lambda \\
 -\rho\frac{ d\Phi_{\rm grav}}{dr}v_r +S_E \\
\end{bmatrix}.
\label{eq:Smatrix}
\end{equation}
In these equations, $\rho$ denotes the rest-mass density of the fluid. The quantities $M_r = \rho v_r$ and $M_{\theta} = \rho v_{\theta}$ represent the radial and polar momentum densities, respectively, while $M_{\lambda} = \rho \lambda$ corresponds to the angular momentum density. The total energy density is given by $E = \rho v^{2}/2 + e$, which includes both kinetic and internal energy contributions. Here, $p$ denotes the gas pressure, $e$ is the internal energy density, and $v_r$ and $v_{\theta}$ are the velocity components along the $r$ and $\theta$ directions, respectively. The specific angular momentum is defined as $\lambda = r v_{\phi}$, and the squared velocity is $v^{2} = v_{r}^{2} + v_{\theta}^{2} + v_{\phi}^{2}$.
The angular momentum equation includes a source term, $S_{\lambda}$, which accounts for viscous transport, while the source term $S_E$ in the energy equation represents viscous dissipation. The explicit forms of $S_{\lambda}$ and $S_E$ are given by,
\begin{equation}
S_{\lambda}= {\frac{1}{r^2}\frac{\partial}{\partial r}\left(r^3 W_{r\phi}\right);} ~
~S_E= \frac{1}{r^2}\frac{\partial ( r^2 v_{\phi} W_{r\phi})}{\partial r},
\label{eq:sesl}
\end{equation}
The quantity \(W_{r\phi}\) denotes the \(r\!-\!\phi\) component of the viscous stress tensor and given by,
\begin{equation}
W_{r\phi} = \eta_v r \frac{d\Omega}{dr}.
\label{eq:stress}
\end{equation}
where, $\Omega$ is the angular velocity, $\eta_v = \rho \nu$ is the dynamic viscosity coefficient, $\nu= (\alpha p)/(\rho \Omega_k)$ is the kinematic viscosity, $\alpha$ is the Shakura–Sunyaev viscosity parameter. $\Omega_{\rm k}$ is the local Keplerian angular velocity. It may be noted that $\alpha$ viscosity is phenomenological, and the physical origin of such a viscosity is hidden inside the $\alpha$ parameter.
Recently, the origin of such viscosity by a variety of processes has been proposed, for e. g.; magnetorotational instability (MRI; \cite{1991ApJ...376..214B}), gravitational instability (GI; \cite{2003MNRAS.339..937G}), the Rayleigh–Taylor instability (\cite{2018MNRAS.478.1837M}), or supernova explosions (\cite{2021ApJ...906...15M}) and Accretion-modified Stars (\cite{2026ApJ..1000...77L}) for the AGN disk, hydrodynamical turbulence \citep{2005ApJ...629..373A,2005ApJ...629..383M} for colder part of the AGN disk. Each one of them or a combination of these processes might be the physical mechanism behind accretion disk viscosity. However, the jury is still out. We retain the phenomenological form of viscosity. Since in this work, angular momentum transport due to shear ($d\Omega/dr$) or temperature ($\Theta$) or the radial velocity ($v_r$) is more important than the exact value of $\alpha$.

An additional closure relation for the thermodynamic variables $e, p, \rho$ is required. We adopt a variable-adiabatic-index equation of state (EoS) for the gas, as proposed by \cite{2009ApJ...694..492C}, commonly known as the CR EoS.  CR EoS has been used in numerical simulations to study various astrophysical problems \citep{2013ASInC...9...13C, 2022MNRAS.509...85J, 2022ApJ...933...75J, 2023ApJ...948...13J, 2024ApJ...971...13J,2025ApJ...979...61T,2025ASSP...61...15C,2026Tripathi_et_al}. The advantage of using CR EoS is that the adiabatic index becomes a function of temperature and the gas composition.
Recently, \cite{2021MNRAS.502.5227J} presented the alternative form of CR EoS, which we adopt in our analysis.
\begin{equation}
 e = \rho c^2 f
 \label{eq:eos}
\end{equation}
textbf{where $f$ is given as,} 
\begin{equation}
 f = 1 + (2-\xi)\Theta\left[\frac{9\Theta+6/\tau}{6\Theta+8/\tau}\right]+\xi\Theta\left[\frac{9\Theta+6/\tau\eta}{6\Theta+8/\tau\eta}\right]
 \label{eq:f}
 \end{equation} 
$\rho$ is given by $\rho=\sum n_i m_i = n_e m_e(2-\xi+\xi/\eta)$, where $\eta = m_e/m_p$ and $\xi = n_p/n_e$. Here, $n_p$ and $n_e$ denote the proton and electron number densities, and $m_p$ and $m_e$ are their respective rest masses. We consider a pure electron–proton $(e^- - p^+)$ flow, hence $\xi = 1$. The dimensionless temperature is defined as $\Theta = p/\rho c^2$, and $\tau = 2-\xi + \xi/\eta$. The specific enthalpy is expressed as,
\begin{equation} 
 h=(e+p)/\rho=(f+\Theta)c^2
 \label{eq:h}
\end{equation}
The polytropic index $N$ of the flow is,
\begin{equation} 
\begin{aligned}
 N=\rho\frac{\partial h}{\partial p}-1=\frac{\partial f}{\partial\Theta} =6\left[(2-\xi)\frac{9\Theta^2+24\Theta/\tau+8/\tau^2}{(6\Theta+8/\tau)^2}\right] \\
 +6\xi\left[\frac{9\Theta^2+24\Theta/(\eta\tau)+8/(\eta\tau)^2}{(6\Theta+8/(\eta\tau))^2}\right]
 \label{eq:N}
 \end{aligned}
\end{equation}
So, the adiabatic index is
\begin{equation}
 \gamma=1+\frac{1}{N}
 \label{eq:adiabatic}
\end{equation}
The sound speed is $c_s=\sqrt{\gamma~\Theta}$.
The steady-state semi-analytical accretion solution is obtained by imposing $\partial/\partial t \equiv \partial/\partial \theta \equiv \partial/\partial \phi \equiv 0$ and $v_\theta = 0$ in equation (\ref{eq:conserve}). We found the steady-state solution by similar method as in our previous paper \cite{2024MNRAS.528.3964D}. 

\subsection{Spectrum Calculation}\label{sec:spectrum}
We compute the radiative spectrum of the accretion flow using an a posteriori approach, considering three cooling processes: bremsstrahlung, synchrotron emission, and inverse Compton scattering. The bremsstrahlung emissivity (in $\mathrm{erg\ cm^{-3}\ s^{-1}}$) is given by \cite{1973blho.conf..343N},

\begin{equation}
    Q_{\rm br} = 1.4 \times 10^{-27} n_{e}^2 \sqrt{T_e} (1+4.4 \times 10^{-10} T_e).
\label{Bremsstrahlung}
\end{equation}
where $n_{e}$ and $T_{e}$ are the electron number density and temperature, respectively. The synchrotron emissivity is
\begin{equation}
    Q_{\rm syn} = \frac{2\pi k_BT_e}{3c^2}\frac{\nu_t^3}{rr_g}.
\label{Synchrotron}
\end{equation}
where $\nu_t$ is the turnover frequency, computed following \citet{10.1046/j.1365-8711.2000.03297.x}. The magnetic field ($B$) is assumed stochastic, with $B^2/8\pi = \beta p$, giving $B=\sqrt{8\pi \beta p}$, where $\beta = 0.01$ is constant.

Synchrotron photons are inverse-Comptonized by thermal electrons, with emissivity
\begin{equation}
    Q_{\rm ic} = \zeta_{\rm syn} Q_{\rm syn}
\label{Comptonization of synchrotron}
\end{equation}
where $\zeta_{\rm syn}$ is
\begin{equation}
    \zeta_{\rm syn} = 3 \varphi(\Theta_e) \left(\frac{x_t}{\Theta_e}\right)^{\alpha_0-1} \left[\Gamma_{\rm inc}(1-\alpha_0, \frac{x_t}{\Theta_e})+\frac{6\Gamma(\alpha_0)P_{\rm sc}}{\Gamma(2\alpha_0+3)}\right],
\label{zeta syn}
\end{equation}
Here, $\Theta_e ={k_BT_e}/({m_ec^2})$, $\varphi(\Theta_e)= (1+4\Theta_e^2)/(1+40\Theta_e^2)$, $x_t={h\nu_t}/(m_ec^2)$, and $\alpha_0=-\ln P_{\rm sc}/\ln A$. The scattering probability is $P_{\rm sc}=1-\exp(-\tau_{\rm es})$, where $\tau_{\rm es}$ is the optical depth, and $A=1+4\Theta_e+(4\Theta_e)^2$ is the mean amplification factor.

To compute the spectrum, we determine the electron temperature by retaining the velocity and density profiles and solving the coupled equations for proton ($d\Theta_p/dr$) and electron ($d\Theta_e/dr$) temperatures. The general relativistic form of these equations is given in \citet{doi:10.1142/S0218271819500378,2020A&A...642A.209S,2023MNRAS.522.3735S} for transonic accretion flows.

\section{Simulation code and Boundary condition}\label{sec:code}
Our simulation employs a second-order accurate total variation diminishing (TVD) scheme, with details described in our previous work \citep{1995ApJ...452..364R,2024MNRAS.528.3964D}. A key feature of our code is the use of $\rho\lambda$ as the conserved variable instead of $\rho v_{\phi}$, ensuring exact conservation of angular momentum in the absence of viscosity. This makes the scheme strictly angular momentum conserving.

The computational domain is discretized following \cite{2011ApJ...728..142L}, with an exponentially increasing grid,
\begin{equation}
\Delta r_i = \Delta r_1 \, \delta^{\,i-1},
\label{eq:grid}
\end{equation}
where $\Delta r_1$ is the first cell size and $\delta$ is the increment factor. We adopt $\Delta r_1 = 0.025$ and $\delta = 1.001$, yielding 4617 radial grid points over $1.62\rg$ to $2500\rg$. Two ghost cells are used at each radial boundary.

At the outer boundary, matter is injected with density normalized to unity, while the radial velocity $v_r$ and temperature $\Theta$ are taken from analytical solutions.
Since we are looking for QPOs in the range of LFQPO or HFQPO, the time scale involved ranges from $0.01$---$1$ secs. Therefore, the instabilities have to be close to the horizon. To simulate such phenomena, a steady boundary condition is advantageous to identify the intrinsic process/processes that can trigger instabilities in the accretion disk. To mimic the near-horizon behaviour, we impose an absorbing inner boundary. Since the horizon location depends on spin, the inner boundary is chosen accordingly: $2.62\rg$ and $2.48\rg$ for $a=0$ and $0.5$, and $1.62\rg$ for $a=0.9$ and $0.95$.
\begin{table}
\centering
\caption{Details of the initial accretion flow parameters.}
\label{tab:injection_setup1}
\begin{tabular}{r c c c c c }
 \hline
 Model & Spin & $\lou$ &$\vo$& $\tho$ & $\epsilon$ \\
    & $\ak$ & & $\times 10^{-3}$&$\times 10^{-4}$& \\
 \hline
 A0L1 & 0.0  & 3.4 &-11.1165&1.66064  & 1.0001 \\
 A0L2 & 0.0  & 3.5  &-11.1089&1.66075   & 1.0001 \\
 A0L3 & 0.0  & 3.6   &-11.1010&1.66086 & 1.0001 \\
 \hline
 A5L1 & 0.5  & 3.1  &-11.1165& 1.65810   & 1.0001 \\
 \hline
 A9L1 & 0.9  & 2.45  &-11.1523& 1.65715   & 1.0001 \\
 A9L2 & 0.9  & 2.5 &-11.1583&1.65815  & 1.0001 \\
 A9L3 & 0.9  & 2.52  &-11.1728&1.65982  & 1.0001 \\
 A9E2 & 0.9  & 2.5  &-12.9953&1.39690  & 1.00005 \\
 \hline
 A95L1 & 0.95 & 2.36  &-11.1808&1.65965  & 1.0001 \\
 \hline
\end{tabular}
\end{table}
\begin{figure*}[!t]
\centering
\includegraphics[width=\textwidth]{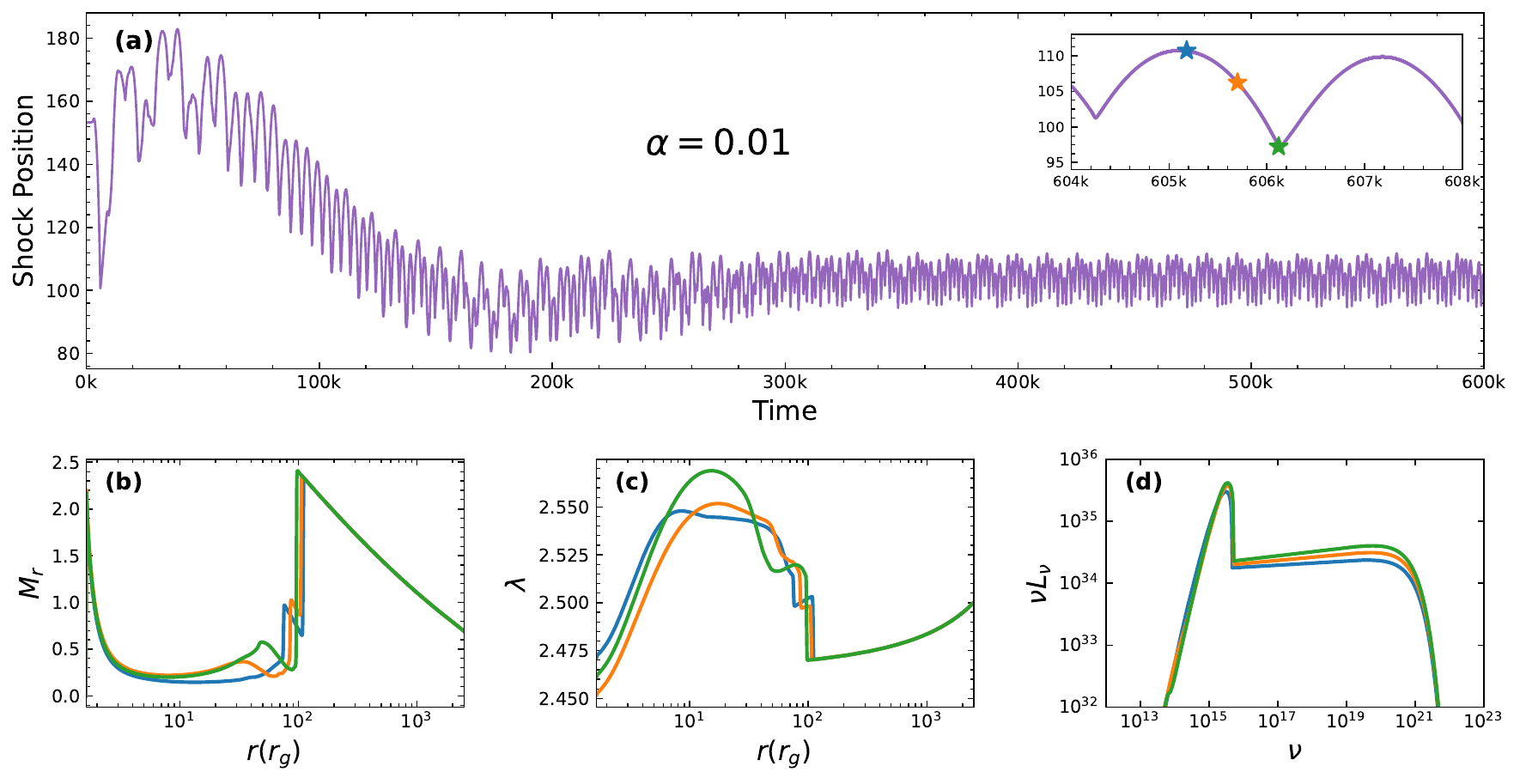}
\caption{\label{fig:1}  Panel (a) shows the complete shock oscillation for $\alpha = $0.01 after adding the viscosity for the model A9L2. Zoomed shock oscillations are shown inside the panel (a), where three points are indicated with different colors. Corresponding Mach number ($M$), angular momentum ($\lambda$), and spectral energy distribution (SED)  of each of the three points are shown in panels (b), (c), and (d), respectively. SED is calculated by adopting a black hole mass of $10\,M_\odot$, an accretion rate of $0.03$ $\dot{M}_{edd}$ ($\dot{M}_{edd}$ is the Eddington accretion rate).}
\end{figure*}
\begin{figure*}[!t]
\centering
\includegraphics[width=\textwidth]{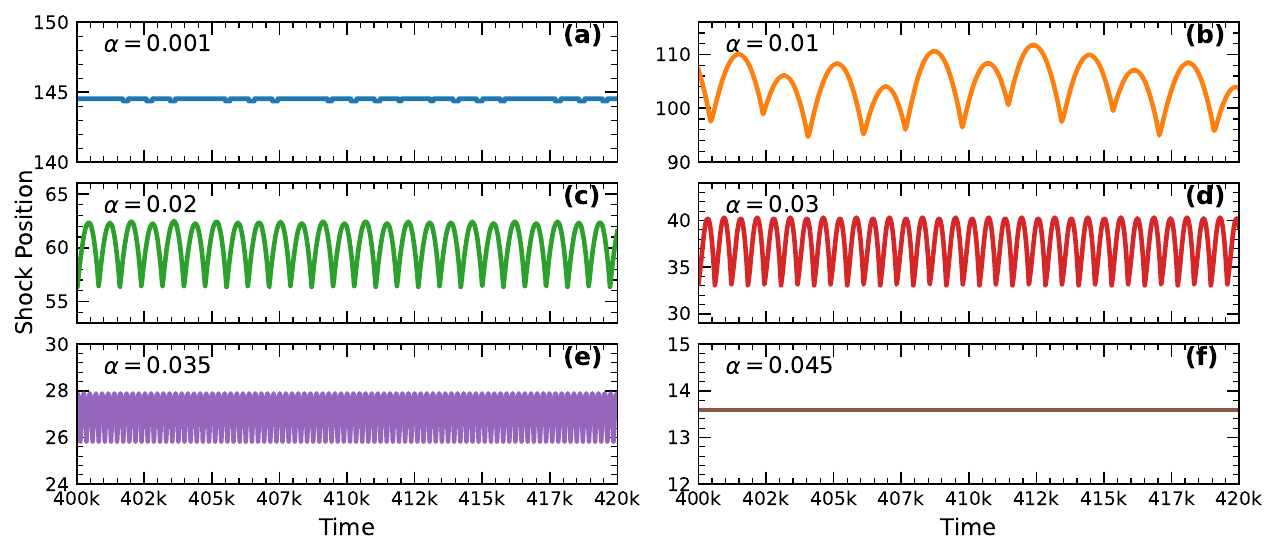}
\caption{\label{fig:2} Panels a, b, c, d, e, f show the zoomed shock oscillation for the time interval [400k to 420k] with $\alpha$=0.001, 0.01, 0.02, 0.03, 0.035, 0.045, respectively. All for the model A9L2.} 
\end{figure*}

\section{Results: Oscillating Shock in Viscous Accretion Flow}\label{results}
Many previous studies \citep{1998MNRAS.299..799L,2011ApJ...728..142L,2012MNRAS.421..666G,2013MNRAS.430.2836G,2014MNRAS.442..251D,2015MNRAS.453..147O,2015MNRAS.448.3221G,2016ApJ...831...33L,2024MNRAS.528.3964D,2025ApJ...994...48D} show that viscosity can trigger shock oscillations and thereby explain the QPOs in black hole candidates. However, all these studies consider a pseudo-potential to mimic a non-spinning black hole. In this study, we use a pseudo-potential to mimic a spinning black hole. We consider different values of black hole spin 0, 0.5, 0.9, and 0.95 as listed in Table \ref{tab:injection_setup1}, to study the effect of spin on the shock oscillation of the accretion flow. The steady-state semi-analytical solutions, which provide the injection values (at the outer boundary $\ro=2500$), the injection speed ($\vo$), and the injection temperature parameter ($\tho$) for the simulation, are inviscid accretion solutions. Specific angular momentum ($\lou$) and the specific energy ($\epsilon$) of the inviscid semi-analytical solutions are also mentioned in Table \ref{tab:injection_setup1}. 

\subsection{Effect of Black Hole Spin}

All three models with spin $\ak=0.0$, i.e., A0L1, A0L2, A0L3 are similar to models L1, L2, and L3 of \cite{2024MNRAS.528.3964D}. These models have the same specific energy ($\epsilon = 1.0001$) but different specific angular momenta ($\lou = 3.4, 3.5, 3.6$). The inviscid shocks appear at 40$r_g$, 103$r_g$, and 564$r_g$, respectively. The viscous effect on the shock oscillations is similar to that reported by \cite{2024MNRAS.528.3964D}: as viscosity increases, the shock moves outward and settles at an outer location, and beyond a critical viscosity parameter ($\alfc$), it begins to oscillate. Even the median shock location increases with viscosity. 
The critical viscosity corresponds to a lower value for higher $\lou$. The values of $\alfc$ for models  A0L1, A0L2, and A0L3 are $0.04,~ 0.02,$ and $0.005$, respectively.
Once the shocks begin oscillating after moving outward, the oscillation timescale becomes long enough that the oscillation frequency is comparatively low. For example, the oscillation frequency ranges from 8.57 Hz to 2.23 Hz for a 10$M_{\odot}$ black hole for the model A0L2.
We also consider a model A5L1 with black hole spin $a_k$ = 0.5. The inviscid shock forms at around 137$r_g$. In this model, shock oscillations begin at $\alfc = 0.01$, while the mean shock position moves outward at around 163$r_g$. The oscillation frequency is 9.5 Hz for a $10 M_{\odot}$ black hole for $\alpha = 0.01$. If we further increase the viscosity, the shock moves even farther outward and oscillates at a lower frequency, similar to the models with $\ak=0$.
 
 For BHs with high spin $\ak=0.9$, we consider three different models A9L1, A9L2, A9L3 having different angular momenta as listed in Table \ref{tab:injection_setup1}. The model A9L1 admits a steady shock solution at $\rs = 32 \rg$ in the inviscid case. We then rerun the simulation in the presence of viscosity, i.e., for $\alpha > 0$.
In these models, the shock moves inward with higher viscosity. For model A9L1, the inviscid shock at 32$\rg$ settles to 26$\rg$ for $\alpha = 0.005$. For $\alpha = 0.01$ and $0.02$, the shock oscillates with mean positions of 22$\rg$ and 17.4$\rg$, respectively, with corresponding frequencies of 109 Hz and 161 Hz. For $\alpha = 0.025$, the shock eventually stabilizes at 13.6$\rg$. Since the shock starts to oscillate after the viscosity is turned on, it is clear that the viscous transport of $\lambda$ has destabilized the shock.
Model A9L2  admits an inviscid stable shock at $\rs = 154 \rg$, which is our representative case.
By varying the viscosity parameter $\alpha$, we then rerun the simulations. 
Figure \ref{fig:1}(a) shows the time variation of shock location with $\alpha=0.01$ for model A9L2. 
The inset shows a zoomed view of the shock oscillation in the time series. The blue, saffron, and green markers indicate the shock locations at three representative phases. The corresponding flow variables, the Mach number $M_r = v_r/c_s$ (Figure~\ref{fig:1}b) and specific angular momentum $\lambda$ (Figure~\ref{fig:1}c), are plotted at these locations. These profiles illustrate the evolution of the solution as the shock moves from its outermost position (blue) through an intermediate state (saffron) to its innermost position (green).
The oscillation significantly perturbs the post-shock region, while the pre-shock flow remains largely unaffected. Consequently, the variability is primarily imprinted on the radiation emerging from the post-shock flow. Since this region dominates the power-law high-energy emission, the shock oscillation naturally leads to QPOs in the hard radiation. Previous studies \citep{2011ApJ...728..142L,2016ApJ...831...33L,2024MNRAS.528.3964D} have shown that the emitted luminosity and shock location are out of phase, with luminosity peaking when the shock is closest to the black hole.
In Figure~\ref{fig:1}d, we present the spectral energy distribution (SED) for model A9L2 corresponding to the three shock locations marked in panel (a).
As the shock attains its maximum extent (blue), the luminosity is lower, and the spectrum is softer. As it moves inward to an intermediate position (saffron), the luminosity increases and the SED becomes harder. When the shock reaches its minimum location (green), the post-shock region is maximally compressed, leading to enhanced luminosity and a harder spectrum.
In Figure~\ref{fig:2}(a–f), we show the temporal evolution of the shock location $\rs$ for different values of $\alpha$ for model A9L2. For $\alpha = 0.001$, the shock settles at $\rs \lesssim 145\rg$ with only weak perturbations. At higher viscosity (i.e., $\alpha \geq \alfc = 0.003$), the shock oscillates. With increasing $\alpha$, the oscillations become more regular, the mean shock location shifts inward, and the oscillation frequency increases. The mean shock positions for $\alpha = 0.01$, 0.02, 0.03, and 0.035 are 100, 57, 35, and 26, respectively. Correspondingly, for a $10M_{\odot}$ black hole, the dominant oscillation frequency increases from 6.09 Hz to 88 Hz as $\alpha$ varies from 0.01 to 0.035. For $\alpha > 0.035$, the shock becomes steady, as shown in Figure~\ref{fig:2}(f).

\begin{figure*}[!t]
\centering
\includegraphics[width=\textwidth]{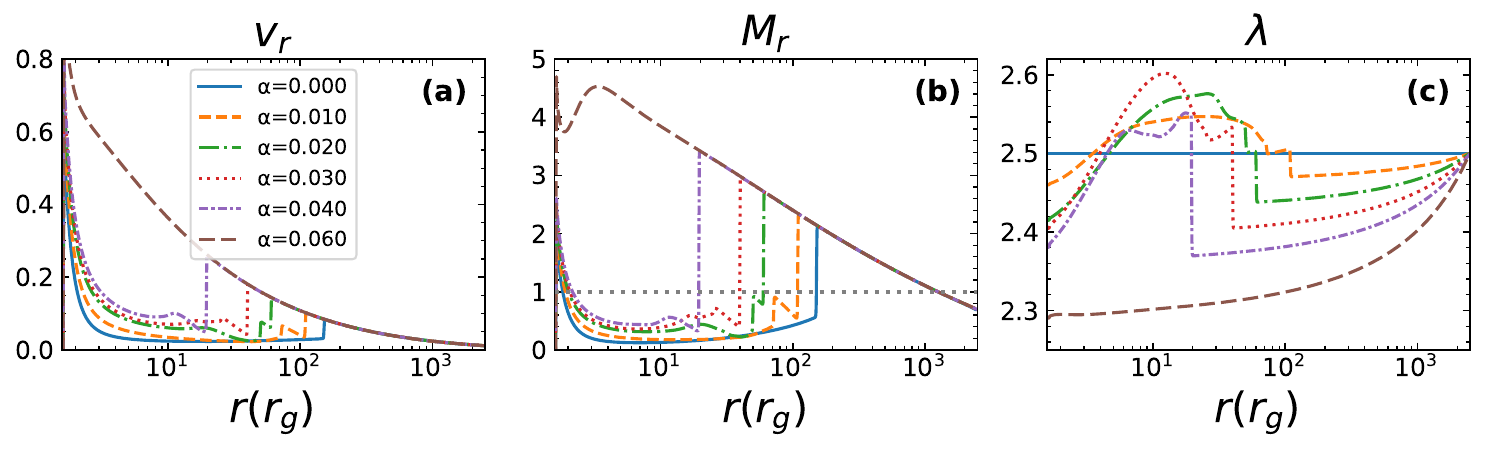}
\caption{\label{fig:3}Flow variables velocity $(v_r)$, Mach number $(M_r)$, specific angular momentum ($\lambda$) for different viscosities are shown in panel (a)-(c) for the model A9L2. The viscosities are shown in the figure.}
\end{figure*}

\begin{figure}[!t]
\centering
\includegraphics[width=0.48\textwidth]{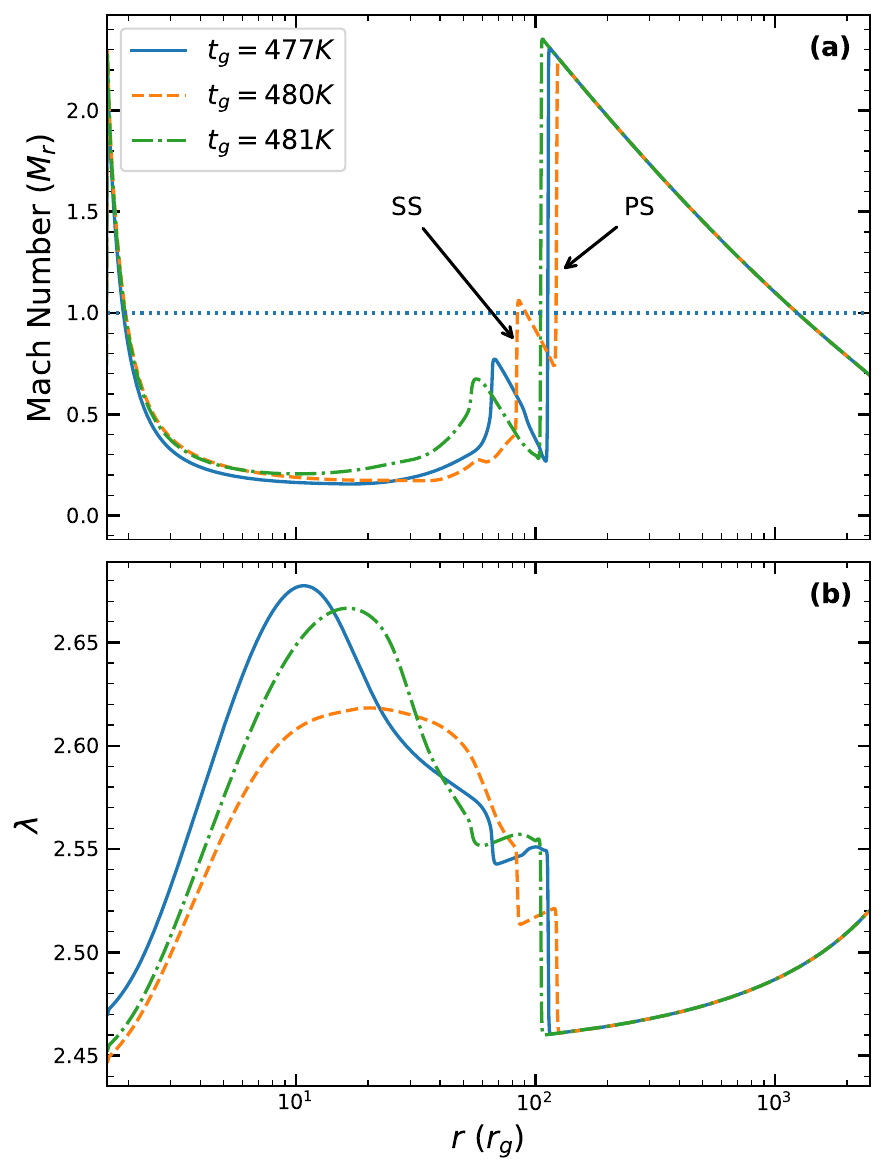}
\caption{\label{fig:ss} Snapshots of the Mach number ($M_r$) and angular momentum ($\lambda$) at different epochs (as indicated in the figure) for model A9L3 are shown in panels (a) and (b), respectively. The viscosity parameter is $\alpha = 0.02$. At $t_g = 480,\mathrm{K}$, both a primary shock (PS) and a secondary shock (SS) are present in the accretion flow and are indicated by the black arrow in panel (a).} 
\end{figure}

In Figure \ref{fig:3}, flow velocity $(v_r)$, Mach number $(M_r)$, specific angular momentum ($\lambda$) are plotted for model A0L2 with $\alpha=0.0,~0.01,~0.02,~0.03,~0.04$, and $0.06$, with colors marked in panel (a). Figure \ref{fig:3} confirms the time variation of the shock front, as shown in Figure \ref{fig:2}. The primary shock (seen as vertical jumps in each variable) shifts inward as viscosity increases, and the post-shock flow wiggles due to the oscillating shock.
In this case, the oscillating amplitude of the shock front is about $\sim 10\rg$.
For $\alpha~ \rightarrow ~ 0.01$--$0.04$, the shock oscillates. However, for higher $\alpha$, such as $\alpha=0.06$ (brown), angular momentum drops enough that the shock disappears, resulting in a Bondi-type solution (one sonic point at a large distance from the BH).

\begin{figure*}[!t]
\centering
\includegraphics[width=\textwidth]{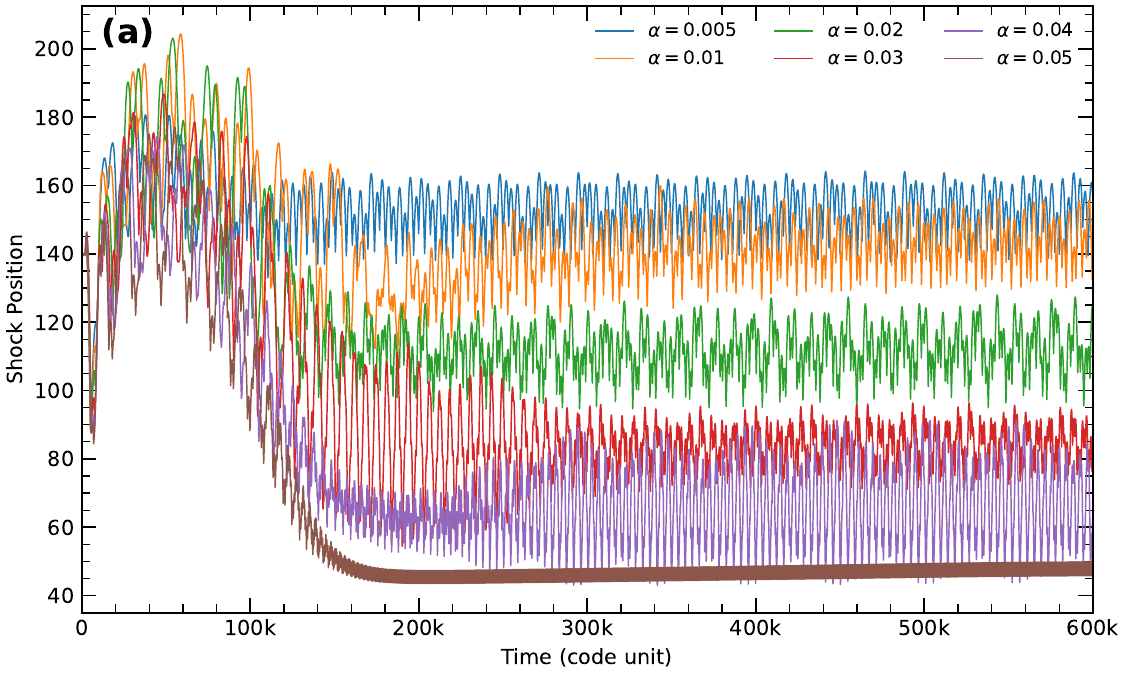}
\caption{\label{fig:shock_pos}Shock position vs time for the models A9E2, as shown in panels (a). The shock is moving inwards with increasing $\alpha$ in this model as well.}
\end{figure*}

 Interestingly, there is a significant difference between our previous papers \citep{2024MNRAS.528.3964D,2025ApJ...994...48D} and the present one. Our earlier studies 
for non-rotating BH ($\ak$ = 0.0), showed that the mean position of an oscillating shock front increases with the increase in $\alpha$. This resulted in a decrease in the oscillation frequency as $\alpha$ increased. As a result, the frequencies were comparatively low, typically sub-hertz to about 10 hertz for a $10M_{\odot}$ black hole. In contrast, in the present study of accretion shock around a highly spinning BH ($\ak=$ 0.9), the mean position of the oscillating shock decreases with increasing $\alpha$, leading to higher oscillation frequencies compared to the non-spinning case.
In model A9L3, the injection angular momentum is increased ($\lou = 2.52$). The inviscid shock is at 677$r_g$. All injection parameters are listed in Table \ref{tab:injection_setup1}.
As viscosity is turned on in A9L3, a secondary shock arises near the black hole \citep[see also][]{2011ApJ...728..142L,2016ApJ...831...33L,2024MNRAS.528.3964D}.
This secondary shock moves outward, and the primary shock moves inward over time; eventually, they merge into a single shock that moves inward and begins to oscillate for low viscosity ($\alpha = 0.005$). Figure~\ref{fig:ss} shows the temporal snapshots of the radial Mach number ($M_r$) and angular momentum ($\lambda$) for this model with $\alpha = 0.02$, during the phase of shock oscillations. In this case, the shock front oscillates between $104\,\rg$ and $127\,\rg$. 
As shown in Figure~\ref{fig:ss}a, at $\tg = 477\,\mathrm{K}$, a primary shock (PS) is located at $123\,\rg$. At $\tg = 480\,\mathrm{K}$, both a primary shock (PS) and a secondary shock (SS) are present at $122\,\rg$ and $83\,\rg$, respectively, shown by the black arrows in Figure~\ref{fig:ss}a. Subsequently, the SS merges with the PS and forms a single primary shock, as shown by the green curve at $\tg = 481\,\mathrm{K}$.    
In this model, shock oscillations persist up to $\alpha$ = 0.05, with a mean shock position at 19$r_g$. In contrast, for $\alpha$ = 0.06, the shock vanishes, and the solution becomes Bondi-like. Therefore, in this case, shock oscillation occurs over a wide range of viscosities, with oscillation frequencies spanning 1.9–152 Hz.
While the critical viscosity required for shock oscillation varies across models, it is important to note that the overall behavior in the presence of viscosity remains qualitatively similar across a wide range of initial shock locations. 
In model A9E2, we considered a flow of less specific energy ($\epsilon$ = 1.00005). The injection parameters are mentioned in Table \ref{tab:injection_setup1}. The evolution of the shock position is shown in Figure \ref{fig:shock_pos} for A9E2. The initial inviscid shock is at 140$\rg$. In this case, the critical viscosity for the shock oscillation is $\alfc=0.003$. The mean position of the shock oscillation is higher than the inviscid shock position (blue and saffron curves) for the lower viscosity range ($0<\alpha \leq 0.01$). But for higher $\alpha$, the mean position of the oscillatory shock moves inwards with the increase in $\alpha$ (green, red, blue, and brown curves in Figure \ref{fig:shock_pos}). The shock oscillates within a frequency range of 1.8 to $80~\mathrm{Hz}$.
 
\begin{figure}[!t]
\centering
\includegraphics[width=0.48\textwidth]{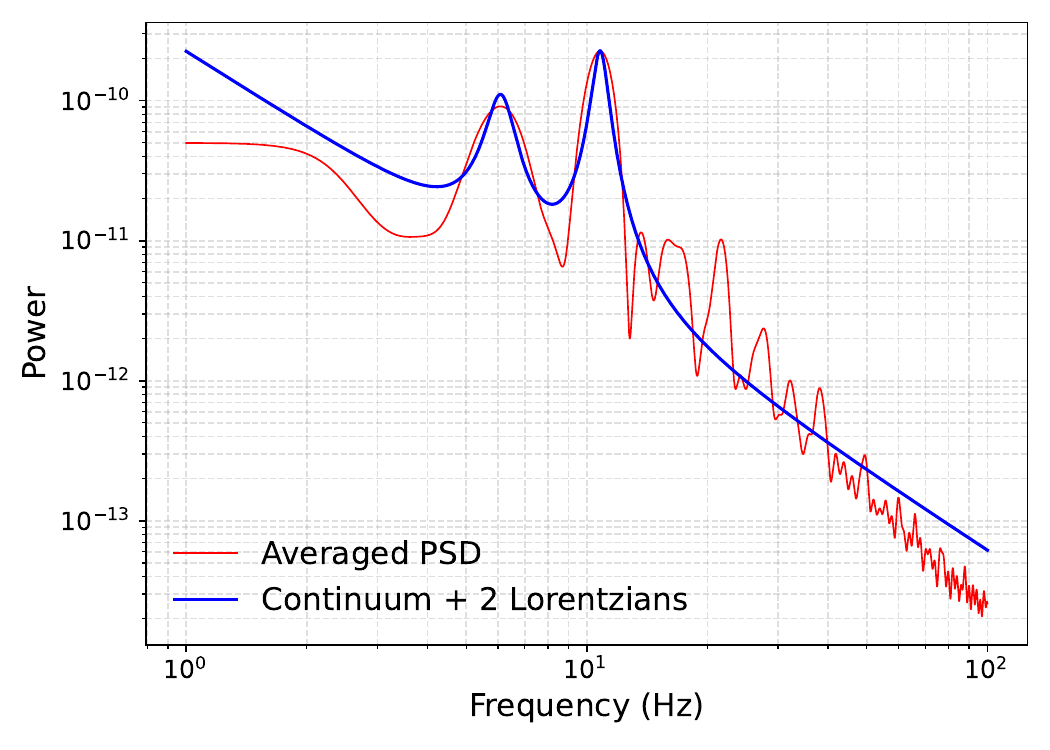}
\caption{\label{fig:psd} Segment averaged Power Density Spectrum (PSD) (in red) and the fitted Lorentzians over a continuum Power Law (in blue) are shown. Corresponding shock oscillation is for the model A9L2 with $\alpha$ = 0.01, as shown in Figure \ref{fig:2}(b).} 
\end{figure}

 For low and moderate spin cases ($a_k = 0.0$ and $0.5$), increasing viscosity systematically shifts the shock front outward. Beyond a critical viscosity, the solution topology changes: the initial multi–sonic–point configuration collapses to a single inner sonic point solution. This trend agrees with earlier studies using the Paczyński–Wiita potential, where shocks 
consistently drift outward. In those works, the computational domain was relatively small with supersonic outer boundary conditions, causing shocks to move outward and eventually exit the domain before the topology transition could be fully followed.
In contrast, the high-spin case ($a_k = 0.9$) shows qualitatively different behaviour. As viscosity increases, the shock moves inward toward the black hole. Beyond a critical value, the shock begins to oscillate due to differential angular momentum transport between the pre- and post-shock regions. The mean shock location ($\langle \rs \rangle$) decreases with increasing $\alpha$ until a second critical viscosity is reached, beyond which the shock disappears, and the flow transitions to a solution passing through a single outer sonic point. This inward migration with viscosity is consistently observed for all models with $\ak = 0.9$.

\begin{figure*}[!t]
\centering
\includegraphics[width=\textwidth]{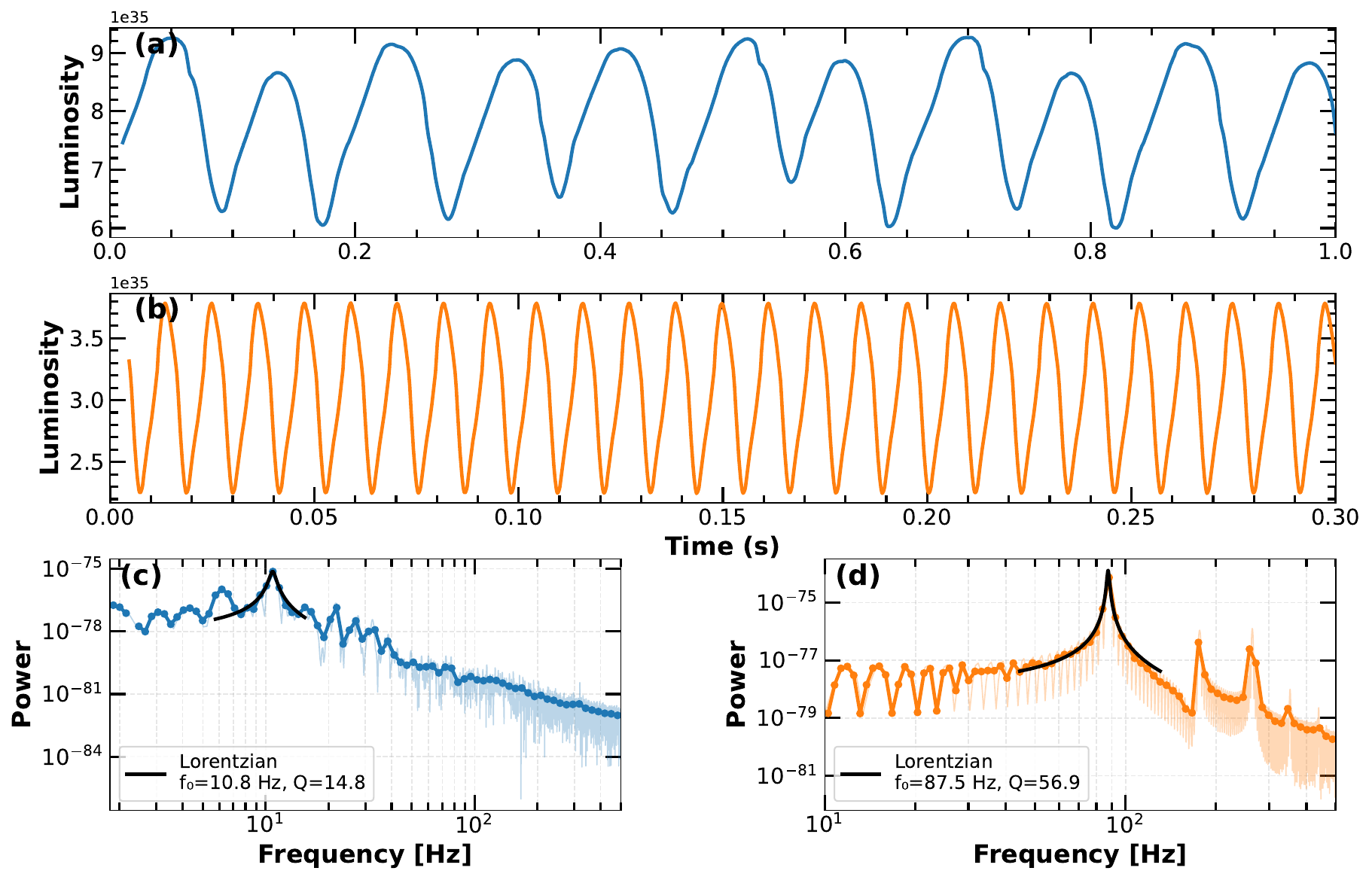}
\caption{\label{fig:5} Time series of luminosity for two different $\alpha$=0.01, 0.035, and their corresponding Power Density Spectrum (PSD) is shown for the model A9L2. We assume black hole mass is 10$M_{\odot}$.} 
\end{figure*}

To determine whether this inwards migration of shock is unique for spin ($\ak$ = 0.9) or is a generic feature of highly spinning black holes, we additionally considered a model A95L1 with a higher spin value of $\ak$ = 0.95, as listed in Table~\ref{tab:injection_setup1}. The inviscid flow exhibits a steady shock at 278$\rg$. Introducing a small viscosity of $\alpha$ = 0.005 destabilizes the shock, causing it to oscillate with a frequency of 2.70 Hz and an amplitude of 16$\rg$, with the mean position shifting inward to 260$\rg$. Thus, in this model as well, even a modest viscosity leads to the shock moving inward toward the black hole.
Shock oscillations persist up to $\alpha$ = 0.055, with the mean position continuously drifting inward and reaching 23$r_g$, and the corresponding oscillation frequency is 71 Hz. Finally, the oscillatory behavior is completely suppressed, and the shock becomes steady at 15$\rg$ for an even higher viscosity ($\alpha$ = 0.06).
Overall, the quantitative behavior observed for the model A95L1 ($\ak$ = 0.95) closely resembles that of the model with $\ak$ = 0.9: viscosity induces an inward migration of the shock, followed by its eventual disappearance at higher viscosities. This suggests that the inward-moving trend is a characteristic feature of rapidly spinning black hole accretion flows, marking a qualitative departure from the behavior observed in models with non-spinning or slowly spinning black holes.

\begin{table}
\centering
\caption{QPO centroid frequencies and quality factors for the model A9L2 and A9E2.}
\label{tab:qpo_prop}
\begin{tabular}{cccccc}
\hline
Model & $\alpha$ & $f_{\rm QPO,1}$ (Hz) & $Q_1$ & $f_{\rm QPO,2}$ (Hz) & $Q_2$ \\
\hline
{A9L2}
 & 0.003  & 12.62   &  25.5  & 11.5  & 45.0  \\
 & 0.004  & 7.22   &   9.7 & 24.31 & 25.9  \\
 & 0.005  & 7.69   &   8.5 & 15.34 & 31.9  \\
 & 0.0075  & 5.14   &  5.1 & 8.84 & 9.1  \\
 & 0.01   & 6.09   & 6.0  & 10.79 & 10.8  \\
 & 0.015   & 9.19   & 9.6  & 16.20 & 16.8  \\
 & 0.02   & 14.04  & 21.7 &  25.26 & 29.7  \\
 & 0.025   & 37.90  & 51.8 &  75.90 & 100.4  \\
 & 0.03   & 32.70  & 43.5   & 65.40 & 76.1  \\
 & 0.035  & 88.22  & 111.3 &  176.48 & 202.1 \\
\hline
{A9E2}
 & 0.003 & 6.28  & 11.7 & 12.57  & 47.2  \\
 & 0.005 & 3.22  & 5.8 & 5.04  & 11.7  \\
 & 0.01  & 3.60  & 6.1 & 6.79  & 13.7  \\
& 0.015  & 4.37  & 7.6 & 8.14  & 14.4  \\
 & 0.02  & 1.70  & 3.0 & 5.69  & 7.7 \\
 & 0.03  & 2.84  & 5.3 & 9.23 & 15.4  \\
 & 0.04  & 4.58  & 7.3  & 13.93 & 26.9   \\
 & 0.05  & 40.60  & 61.1 & 81.12 & 117.5  \\
 & 0.06  & 12.11  & 23.8 & 34.12 & 33.2  \\
  & 0.065 & 78.15  & 15.5 & 157.88 & 13.0 \\
\hline
\end{tabular}
\end{table}

\subsection{Power Density Spectra (PDS) of Shock Oscillations}
A segment-averaged power density spectrum (PDS) is constructed from the simulated shock oscillations by selecting a fixed time interval. The signal is divided into equal-duration segments with a chosen fixed fractional overlap. A Lomb-Scargle periodogram is computed on a common frequency grid for each segment. Each periodogram is averaged to reduce the variance of the raw spectral. The broadband variability is described by a power-law continuum, \(P_{\rm cont}(f) = C_{0} f^{-\alpha_k}\), which is fitted globally while excluding the frequency range dominated by quasi-periodic oscillations (QPOs). Within these combined QPO bands, the PDS is modeled as a composite function comprising a fixed continuum and two Lorentzian components, each specified by a centroid frequency and a half-width at half-maximum (HWHM), thereby capturing two distinct QPO features. To avoid spuriously narrow, unresolved peaks, the Lorentzian widths were constrained to be larger than a floor set by the inverse of the total time baseline. Figure~\ref{fig:psd} shows a representative segment averaged power density spectrum (PDS; red curve) together with the best-fitting model, consisting of two Lorentzian components superposed on a power-law continuum (blue curve).
The PDS corresponds to the shock oscillation for the model A9L2 with \(\alpha = 0.01\), as illustrated in Figure~\ref{fig:2}(b), assuming a black hole mass of \(10M_{\odot}\). For this analysis, a fixed time interval of \(t = 15\)-\(30~\mathrm{s}\) is selected, and the synthetic light curve is divided into segments of duration \(0.5~\mathrm{s}\) with a fractional overlap of
0.5. Although the averaged PDS exhibits several peaks, only two Lorentzian components were fitted on top of the continuum. The continuum fit yields \(C_{0} = 2.2\times10^{-10}\) and a slope \(\alpha_{k} = 1.82\). The two Lorentzians have centroid frequencies of \(6.09~\mathrm{Hz}\) and \(10.79~\mathrm{Hz}\), with corresponding quality factors \(Q \simeq 6\) and \(Q \simeq 10\), respectively. The same procedure is applied to the shock oscillations obtained for the models, A9L2 and A9E2, with different values
of \(\alpha\), and the resulting peak frequencies ($f_{\rm QPO,1}, f_{\rm QPO,2}$) and the corresponding quality factors ($Q_1, Q_2$) of both Lorentzians are summarized in Table~\ref{tab:qpo_prop}. It may be noted that the QPO frequency for both models first decreases with the increase of $\alpha$ and then increases with $\alpha$. Although the median shock location is decreasing with the increasing values of $\alpha$, for low values of $\alpha$, the amplitude of oscillation is smaller; therefore, the oscillation frequency decreases. For higher values of $\alpha$, a similar trend is observed: although the mean shock location decreases with increasing viscosity, the oscillation amplitude also influences the resulting frequency. Consequently, the QPO frequency does not exhibit a strictly monotonic dependence on the viscosity parameter, as summarized in Table~\ref{tab:qpo_prop}.

\begin{figure}[!t]
\centering
\includegraphics[width=0.5\textwidth]{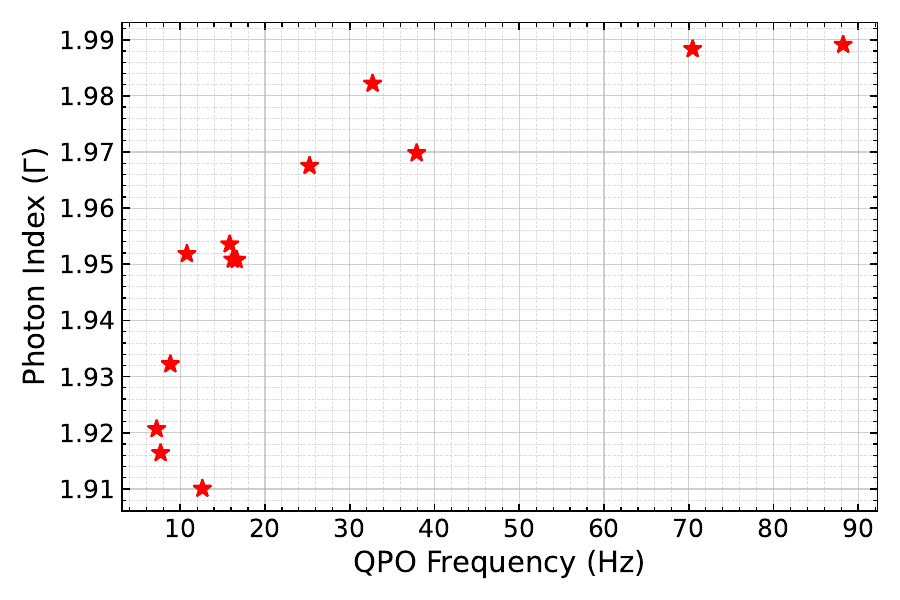}
\caption{\label{fig:7} Photon index vs QPO centroid frequency for the models A9L2 (considering black hole mass is 10 $M_\odot$). } 
\end{figure}

\subsection{Implications for the Black Hole System}
To connect the shock oscillation with the observable timing features, we compute the synthetic luminosity light curve for the A9L2 model. In Figure~\ref{fig:5}a, and \ref{fig:5}b, we present the time series of luminosity for two representative viscosity parameters, $\alpha$ = 0.01 and 0.035, assuming a black hole mass of 10 $M_\odot$. In both cases, the luminosity exhibits a clear quasi-periodic modulation which directly reflects the periodic compression and rarefaction of the post-shock region, whose thermal and dynamical properties regulate the emitted radiation.
The corresponding power density spectra (PDS) are shown in the lower panels of Figure~\ref{fig:5}c, and \ref{fig:5}d, which reveal distinct peaks marking the dominant oscillation frequencies. We have fitted the Lorentzian to the peak frequency for these two cases. These PDS exhibit narrow QPO-like peaks whose centroid frequency and coherence increase with viscosity: for $\alpha=0.01$ the dominant feature occurs at $\nu_0\simeq 10.8\,\mathrm{Hz}$ with $Q\simeq 14$, while for $\alpha=0.035$ it shifts to $\nu_0\simeq 88\,\mathrm{Hz}$ with $Q\simeq 57$. These frequencies match the shock-oscillation frequencies obtained from the radial shock-position analysis. This agreement confirms that the dynamical oscillation of the shock front indeed drives the luminosity variability.
In observational timing work, the coherence is commonly quantified by the quality factor $Q \equiv \nu_0/\mathrm{FWHM}$. Black-hole X-ray binaries frequently show strong low-frequency QPOs whose centroid drifts over a broad range during outbursts \citep{2019NewAR..8501524I,2024ApJ...963..118W,2025MNRAS.540.1394B}.
In particular, recent broadband spectral-timing analyses of the bright 2023 outburst of \textit{Swift}~J1727.8$-$1613 report prominent type-C QPOs with rapidly evolving centroid frequency and coherence, highlighting that multiple Lorentzian components and frequency migration are often required to describe real PDSs \citep{2025MNRAS.540.1394B}.
At the same time, highly coherent low-frequency QPOs with $Q\gtrsim 50$ have also been reported in the ``heartbeat'' black-hole system IGR~J17091$-$3624 \citep{2024ApJ...963..118W}.
The high-frequency QPOs in GRS~1915+105 show a prominent peak at $\sim$67\, Hz, and a second HFQPO feature at a higher frequency has also been reported in the same source \citep{1997ApJ...482..993M,2001ApJ...554L.169S}.
More recently, renewed timing analyses confirm that the HFQPO in GRS~1915+105 persists around $\sim$65--71\, Hz in specific variability classes, strengthening its status as a long-lived HFQPO phenomenon \citep{2025MNRAS.540...37M,2025PASA...42..144M,2025MNRAS.540.2965H}.
Our analysis demonstrates that viscous transonic accretion flows can sustain shock oscillations over a broad range of frequencies.
PDS analysis reveals multiple Lorentzian components, indicating complex temporal variability. The strong dynamical coupling between shock motion and radiative emission suggests oscillating shocks as a robust physical mechanism for low- to high-frequency quasi-periodic oscillations (QPOs) in systems of rapidly rotating BHs.

Figure~\ref{fig:7} presents how the photon index ($\Gamma$) depends on the QPO centroid frequency ($\nu_{\rm QPO}$) for model A9L2. The photon index is estimated from time-averaged spectra, obtained by averaging over three snapshots spanning one oscillation cycle, for viscous runs with different $\alpha$ that yield distinct QPO frequencies. The spectra are computed following the procedure described in Section~\ref{sec:spectrum}, assuming a black hole mass of $10M_\odot$ and an accretion rate of $0.03\dot{M}_{\rm edd}$.
Observational studies of black hole systems have established a correlation between low-frequency QPOs and the power-law photon index, commonly referred to as the $\Gamma$–$\nu_{\rm QPO}$ relation \citep{2004ApJ...612..988T,2005ApJ...626..298T,2006ApJ...643.1098S,2007ApJ...663..445S}. This correlation has also been employed for black hole mass estimation (e.g., Cygnus~X-1; \citealt{2007ApJ...663..445S}) and has been reported for sources such as MAXI J1535–571 and H~1743–322 \citep{2018ApJ...868...71S,2018AstL...44..378M,2017ApJ...834...88M}. Within the Two-component Advective Flow (TCAF) framework, its physical origin has been explored primarily in the low-frequency regime \citep{2019ApJ...875....4S,2024JHEAp..43...79E,2025NatSR..1511556E}.
In contrast, Figure~\ref{fig:7} covers a broader frequency range. We find a qualitatively similar trend: $\Gamma$ increases with $\nu_{\rm QPO}$ and approaches an approximately saturated value at the high-frequency. However, the relation is not strictly monotonic, as also indicated in Table~\ref{tab:qpo_prop}, where the QPO frequency shows a non-monotonic dependence on $\alpha$, influenced by the outer boundary conditions. The $\Gamma$–$\nu_{\rm QPO}$ distribution exhibits scatter rather than a single-valued relation, consistent with observations. This behavior likely reflects the non-linear response of the post-shock region, which governs the high-energy emission. These results support a shock oscillation origin of QPOs and suggest that the $\Gamma$–$\nu_{\rm QPO}$ correlation extends beyond the low-frequency regime into the high-frequency domain, providing a testable prediction for future observations.

\section{Summary and Discussion}\label{discussion}

In this study, we have examined the viscous, transonic, low-angular-momentum accretion flows around rotating black holes, focusing on the formation, stability, and oscillations of centrifugal pressure-supported shocks.
We have used a pseudo-potential to mimic a Kerr BH and not the exact GR equations of motion. The success of a Kerr pseudo-potential does not imply regenerating every aspect of the Kerr metric, but rather regenerating broad and important features obtained in general relativity. In that respect, the presence of the spin-angular momentum coupling term makes this potential a special one. \cite{2018PhRvD..98h3004D} showed that energy-angular momentum parameter space deviates from the exact GR one by 6-12\%. While \cite{2024IJMPD..3350059B} showed the frame dragging effect using the effective potential. Therefore, it is reasonable to use this potential in the simulation.
In addition, we have used the one-dimensional simulation as seen in equation \ref{eq:conserve}. Although this is a simplifying assumption, it is dictated by the need to go high in resolution and very long runs. This choice of high resolution is made in order to check the robustness of the results. Spurious instabilities tend to exacerbate with higher resolution or longer runs. It may also be noted that the viscosity-induced shock oscillation depends on angular momentum transport in the $r$ direction. So such an approximation does not compromise the generic nature of the results obtained. Although, needless to say, a multi-dimensional simulation would affect the details of the result. However, for multi-dimension, one may have to compromise on resolution.

For non-rotating and slowly rotating black holes ($\ak$ = 0) and ($\ak$ = 0.5), we reproduced results similar to previous work \citep{2011ApJ...728..142L,2013MNRAS.430.2836G,2014MNRAS.442..251D,2015MNRAS.453..147O,2016ApJ...831...33L,2024MNRAS.528.3964D,2025ApJ...994...48D}. An increase in the viscosity parameter, piles up the angular momentum in the post shock disk, strengthens the centrifugal barrier, and consequently pushes the shock front outward. Beyond a critical viscosity, the shock starts to oscillate with the median shock position shifting outward. Beyond another high critical viscosity, the shock solution collapses to a solution passing through a single inner sonic point, eliminating the shock. These results are consistent with earlier simulations using the Paczyński–Wiita potential.
The shock dynamics differ for rapidly rotating black holes ($\ak$ = 0.9, 0.95). In all models, the shock moves inward to the black hole as viscosity increases, and shock oscillations occur over a wide range of $\alpha$. So, in case of high spin, the oscillation frequencies span a wide range from a few Hz up to more than 100 Hz for a 10$M_\odot$ black hole. The inward motion of the shock naturally leads to higher oscillation frequencies because the dynamical timescale becomes lower at a smaller radius.
A physically consistent explanation for these spin-dependent shock behaviors arises from the properties of Kerr BH and from the viscous angular momentum transport. 
High values of BH spin ($\ak \gtrsim 0.9$) imply that the length scales closer to the BH decrease, which causes the flow to be hotter closer to the BH. Hotter flow transports angular momentum more efficiently. The presence of the BH spin and the flow angular momentum coupling term (the term $\propto~\ak \lambda$ in equation \ref{eq:potential}) also complicates the dynamics. As a result, the median shock will form closer to the horizon,
compared to a BH with zero or low spin. As the median shock approaches the BH horizon, it is expected that the shock-oscillation frequency will increase. However, since these are fluid systems, the oscillation amplitude may not follow a linear trend, and the frequency may exhibit non-monotonic behavior.

In this work, we have explained QPOs in a shocked transonic disk by destabilizing the steady shock through viscous redistribution of the flow's angular momentum. While mass accretion rates are also important, they are not considered in this paper, as our focus is on the role of viscosity in determining QPO frequencies.
Building on this, our results indicate that the qualitative behavior of shock motion is strongly dependent on the spin of the black hole. In low-spin systems, viscosity causes the mean shock to migrate outward during shock oscillation. High-spin systems exhibit inward migration of the mean shock. This provides a straightforward physical explanation for why rapidly spinning black holes in X-ray binaries are better able to produce high-frequency QPOs. However, the actual QPO frequency will depend on plasma flow parameters such as injection speed, temperature, angular momentum, and the flow viscosity.
We have also examined the correlation between the QPO frequency and the power-law photon index ($\Gamma$–$\nu_{\rm QPO}$) by computing the spectrum for a $10M_\odot$ black hole.

\section*{Acknowledgements}
We acknowledge ARIES, Nainital, India, for providing computational resources used in this work. 

\appendix
\section{2D Simulation of Viscous Accretion Flow}\label{appendix}
\begin{figure}[!t]
\centering
\includegraphics[width=.48\textwidth]{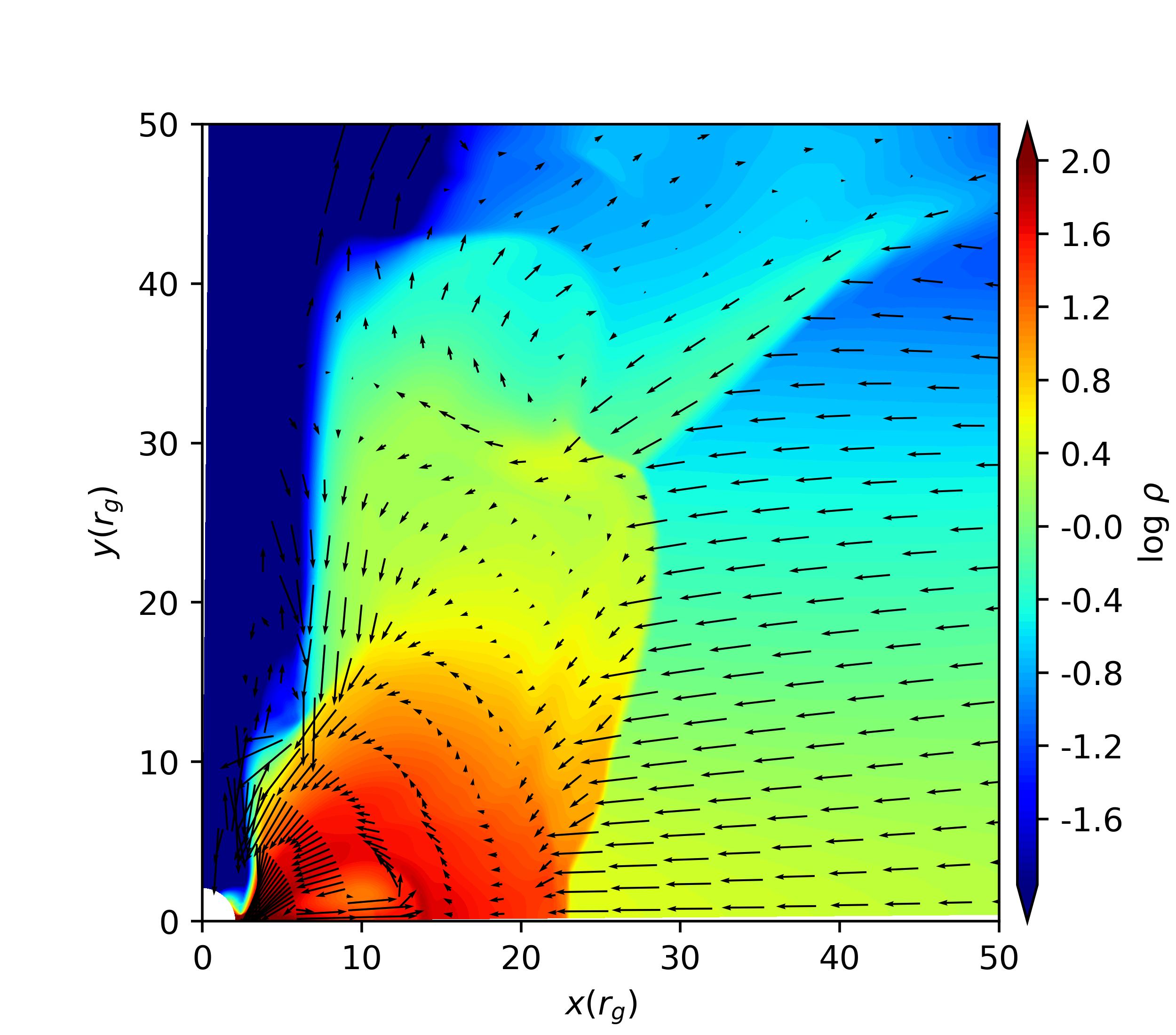}
\caption{\label{fig:3a} Density contour of the inner part of the accretion disc overlaid with the velocity vector at dynamical time t = 15000$t_g$. The spin of the black hole $a_k$ =0.9, viscosity parameter,$\alpha$ = 0.01. Specific energy ($\epsilon$) and specific angular momentum ($\lambda$) at the outer boundary are 1.0005 and 2.4, respectively.}
\end{figure}

\begin{figure}[!t]
\centering
\includegraphics[width=0.48\textwidth]{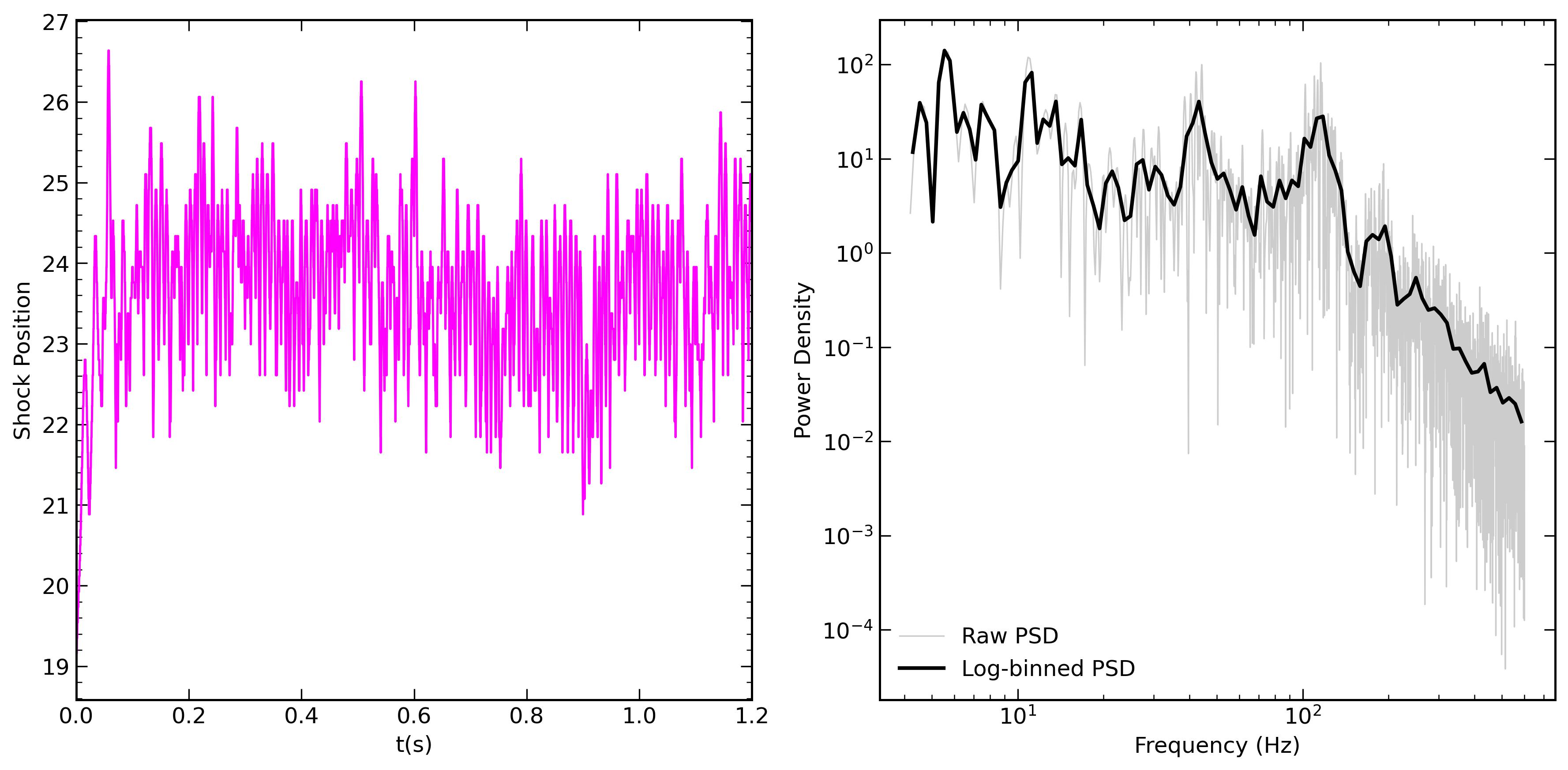}
\caption{\label{fig:4a}Time evolution of shock position ( in left) and the corresponding power density spectrum (PSD) (in right), assuming a $10M_\odot$ black hole. PSD showing oscillation frequency of 42 and 115 Hz.  }
\end{figure}

 {We present a 2D simulation of accretion flows around a black hole with spin $a_k=0.9$.
Figure \ref{fig:3a} shows the density contour (blue-black to red-brown) of the inner region of the accretion flow. Arrows are the velocity vectors, and the red-brown region is the post-shock disk. The shock front is clearly defined
(jump from greenish yellow to red). The corresponding shock oscillation and PDS (power density spectrum) of the resulting radiation are also shown in Fig \ref{fig:4a}. From the PDS, it is clear that shock oscillation can exhibit high frequencies (42 and 115 Hz) in the 2D case as well.}

\bibliographystyle{elsarticle-harv} 
\bibliography{biblio}






\end{document}